\documentclass{aa} 

\usepackage{graphicx}
\usepackage{txfonts}
\usepackage{color}
\usepackage{mathtools}
\usepackage{amsmath}
\usepackage{xcolor}
\usepackage{svg}
\usepackage[colorlinks, linkcolor=blue, anchorcolor=blue, citecolor=blue]{hyperref}

\usepackage{natbib}
\bibpunct{(}{)}{;}{a}{}{,} 

\begin{document}

   \title{A semi-analytical thermal model for craters with application to the crater-induced YORP effect}


   \author{Wen-Han Zhou
   \inst{1}
   \and
   Patrick Michel
   \inst{1,2}
   } 

   \institute{Universit\'e C\^ote d'Azur, Observatoire de la C\^ote d'Azur, CNRS, Laboratoire Lagrange, 96 Bd de l'Observatoire, Nice 06304, France\\
              \email{wenhan.zhou@oca.eu}
        \and
             University of Tokyo, Department of Systems Innovation, School of Engineering, Tokyo, Japan
             }

    \authorrunning{Zhou and Michel}

  \abstract
    {The YORP effect is the thermal torque generated by radiation from the surface of an asteroid. The effect is sensitive to surface topology, including small-scale roughness, boulders, and craters.}
    {The aim of this paper is to develop a computationally efficient semi-analytical model for the crater-induced YORP (CYORP) effect that can be used to investigate the functional dependence of this effect.}
    {This study linearizes the thermal radiation term as a function of the temperature in the boundary condition of the heat conductivity, and obtains the temperature field in a crater over a rotational period in the form of a Fourier series, accounting for the effects of self-sheltering, self-radiation, and self-scattering. By comparison with a numerical model, we find that this semi-analytical model for the CYORP effect works well for $K>0.1 \rm W/m/K$. This semi-analytical model is computationally three-orders-of-magnitude more efficient than the numerical approach.}
    {We obtain the temperature field of a crater, accounting for the thermal inertia, crater shape, and crater location. We then find that the CYORP effect is negligible when the depth-to-diameter ratio is smaller than 0.05. In this case, it is reasonable to assume a convex shape for YORP calculations. Varying the thermal conductivity yields a consistent value of approximately 0.01 for the spin component of the CYORP coefficient, while the obliquity component is inversely related to thermal inertia, declining from 0.004 in basalt to 0.001 in metal. The CYORP spin component peaks at an obliquity of $0^\circ$, $90^\circ$, or $180^\circ$,  while the obliquity component peaks at an obliquity of around $45^\circ$ or $135^\circ$. For a z-axis symmetric shape, the CYORP spin component vanishes, while the obliquity component persists. Our model confirms that the total YORP torque is damped by a few tens of percent by uniformly distributed small-scale surface roughness. Furthermore, for the first time, we calculate the change in the YORP torque at each impact on the surface of an asteroid  explicitly and compute the resulting stochastic spin evolution  more precisely. }
    {This study shows that the CYORP effect due to small-scale surface roughness and impact craters is significant during the history of asteroids. The semi-analytical method that we developed, which benefits from fast computation, offers new perspectives for future investigations of the YORP modeling of real asteroids and for the complete rotational and orbital evolution of asteroids accounting for collisions. Future research employing our CYORP model may explore the implications of space-varying roughness distribution, roughness in binary systems, and the development of a comprehensive rotational evolution model for asteroid groups.}
    

   \maketitle
%

\section{Introduction}

The Yarkovsky-O'Keefe-Radzievskii-Paddack (YORP) effect is a thermal torque that can alter the spin state of an asteroid over time \citep{Rubincam00, Vokrouhlicky02, Vokrouhlicky2015}, and is caused by the asymmetric re-emission of solar radiation by the irregular surface of the asteroid, resulting in a net torque that can spin up or spin down the asteroid's rotation. {{The absorption of solar radiation makes no contributions to the YORP torque as it is averaged out over the spin and orbital periods for any asteroid shapes \citep{Nesvorny2008}.}} So far, 11 asteroids showing time-varying rotational periods  have been detected \citep{Lowry2007, Taylor2007, Durech2022, Tian2022}.

The YORP effect has important implications for an asteroid's long-term rotational evolution. This effect can either spin down the asteroid to an extremely slow rotation, triggering a tumbling motion \citep{Pravec2005}, or spin up the asteroid to its rotation limit (e.g. rotational period of 2.2 hours for rubble pile asteroids), leading to resurfacing \citep{Sanchez2020} and rotation disruption \citep{Scheeres2007b, Fatka2020, Veras2020}. YORP-induced rotational disruption is supported by the observed asteroid pairs \citep{Vokrouhlicky2008,Polishook2014} and binary asteroids \citep{Jacobson2011, Delbo2011, Jacobson2013, Jacobson2016}, including contact-binary asteroids \citep{Rozek2019, Zegmott2021} and binary comets \citep{Agarwal2020}, which evolve under tidal effects and the binary YORP (BYORP) effect after the binary system is formed \citep{Cuk2005, Steinberg2011}. Further potential observational evidence is the abnormal spin distribution and the obliquity distribution of near-Earth asteroids \citep{Vokrouhlicky2003,Pravec2008, Rozitis2013b, Lupishko2014} and main belt asteroids \citep{Lupishko2019}, {although a recent study points out that collisions might reproduce the observed distribution without the involvement of the YORP effect \citep{Holsapple2022}}. The YORP effect can influence the orbital evolution through the Yarkovsky effect, which is a thermal force that depends on the rotational state of the asteroid \citep{Vokrouhlicky2000,Bottke06}. Therefore, understanding the YORP effect is important for correctly estimating the ages of asteroid families based on how much time is needed for the Yarkovsky effect to  cause the observed orbital dispersion of their members from the original orbit following the disruption of their parent bodies \citep{Vokrouhlicky2006,Cuk2015, Carruba2014, Carruba2015, Carruba2016, Lowry2020, Marzari2020}.

\begin{figure*}
    \centering
    \includegraphics[width = \textwidth]{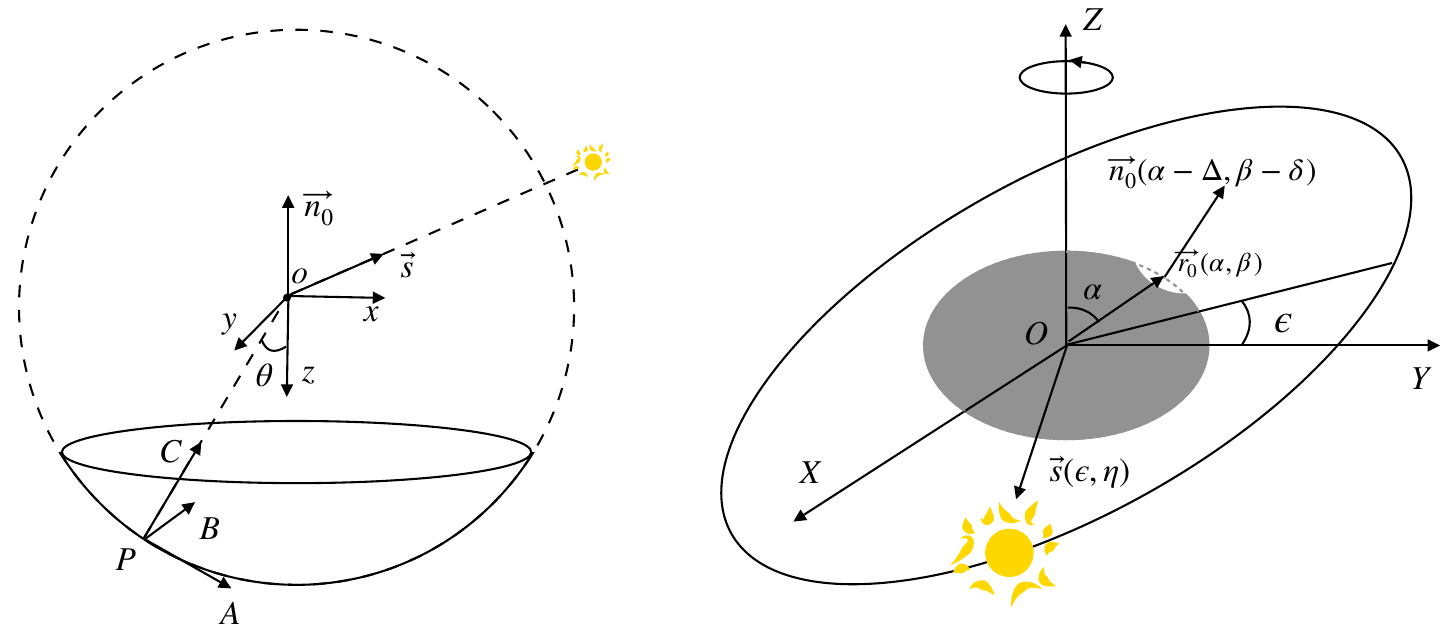}
    \caption{Three coordinate systems in this paper: Coordinate system $oxyz$ for calculating the illuminated domain in the crater, $PABC$ for calculating the effective radiative force of an arbitrary surface element, and $OXYZ$ for averaging the YORP torque over the spin and orbital motion. }
    \label{fig:coordinates}
\end{figure*}

However, accurately calculating the YORP effect on a real asteroid remains a challenge, as it has been demonstrated to be highly sensitive to surface topology \citep{Statler2009, Breiter2009}, such as uniform small-scale roughness \citep{Rozitis2012}, boulders \citep{Golubov12, Golubov14, Sevevcek2015, Golubov17, Golubov2019, Golubov2021, Golubov2022}, and craters \citep{Zhou2022}. Although the YORP torque caused by boulders and the tangential radiative force has been well studied, the YORP effect caused by concave structures has not yet been fully explored. More specifically, there are two kinds of concave structures on asteroid surfaces. The first one corresponds to craters, which result from impacts that an asteroid's surface undergoes during its history and which can cover a large size range and be distributed in various ways. The second corresponds to surface roughness, which corresponds to uniformly distributed small-scale concave features that originate from the continuous effect of various processes that take place at the surface, such as thermal fatigue and space weathering. In our study, we consider both craters and surface roughness.  While a pioneering study by \citet{Rozitis2012} used numerical simulations to investigate the effect of roughness, the high computational expense of such simulations prevents a comprehensive exploration of the functional dependence of this effect and its application to the global spin evolution of asteroid populations.

In addition to the precise calculation of the complete YORP effect, the long-term evolution of asteroids needs to account for stochastic collisions that affect this evolution, because each collision introduces a YORP torque due to the resulting crater. \citet{Bottke2015} performed a first study of the YORP effect accounting for craters caused by collisions, finding strong implications in the age estimate of asteroid families. However, their introduction of the concept of the stochastic YORP effect due to collisions assumed an arbitrary reset timescale for the YORP cycle. In reality, this reset timescale depends on the actual occurrence of each impact causing a crater on the asteroid's surface and the resulting change in the YORP effect. In summary, collisions and the YORP and Yarkovsky effects are all coupled with each other in a way that is so far not well understood.

To account explicitly for the YORP torque caused by roughness and craters, \citet{Zhou2022} developed a semi-analytical model that is computationally efficient for the crater-induced YORP (CYORP) torque. The CYORP torque is defined as the torque difference between the crater and the flat ground:
\begin{equation}
\label{eq:T_CYORP1}
    \vec T_{\rm CYORP} = \vec T_{\rm crater} - \vec T_{\rm ground}.
\end{equation}
In general, it takes the form of the following scaling rule with the radius of the crater $R_0$ and of the asteroid radius $R_{\rm ast}$:
\begin{equation}
    \vec T_{\rm CYORP} = W\frac{\Phi}{c}R_0^2  R_{\rm ast},
\end{equation}
where {{$\Phi$ is the flux of solar radiation at the asteroid's semimajor axis, and  $W$ is the CYORP coefficient, which in turn depends on the obliquity and irregularity of the asteroid, and the depth-to-diameter ratio, location, thermal inertia, and albedo of the crater.}} As a first step, this model assumed a zero thermal conductivity. \citet{Zhou2022} found that roughness or craters that cover 10\% of the asteroid's surface area could produce a CYORP torque comparable to the normal YORP (NYORP) torque, which arises from the macroscopic shape, with ignorance of the fine surface structure. Based on this number, which assumes that all craters have a depth-to-diameter ratio of 0.16, the reset of the YORP torque by the CYORP torque could be as short as 0.4 Myr.

However, the effects of finite thermal conductivity, self-radiation, and self-scattering were not considered in \citet{Zhou2022}. In the present paper, we propose a semi-analytical model that accounts for the effects of self-sheltering, self-radiation, self-scattering, and non-zero thermal conductivity. This new model allows a more general exploration of the functional dependence of the CYORP effect. Moreover, as this semi-analytical model is much more efficient than a purely numerical one, it can be used to study the combined influence of the YORP effect, collisions, and the Yarkovsky effect by incorporating the CYORP effect into the standard evolution model of asteroid families. {{We assume that the craters considered in this work are on a convex asteroid. It is possible to apply the model derived in this work to a moderately concave asteroid by approximating concave structures as craters, but this is beyond the scope of this paper.}}

In this paper, we describe our analytical model for the temperature field in a crater in Section~\ref{sec2}. In Section~\ref{sec3}, we introduce the numerical model that we developed to validate the analytical model. The main results are shown in Section~\ref{sec4} and Section \ref{sec5}. In Section~\ref{sec4}, we analyse the effects of self-sheltering, self-radiation, and self-scattering of the crater, and in Section~\ref{sec5} we show the results of the calculation of the CYORP torque as a function of different variables. For the purpose of illustration, in Section~\ref{sec6} we provide an example of the analysis of CYORP considering a specific real asteroid shape and its surface roughness as well as the consequential rotational evolution. In Section~\ref{sec7} we summarise our main findings and draw conclusions.

\section{Analytical model}
\label{sec2}

\subsection{Calculation of temperature distribution in a crater}

\subsubsection{Linearized analytical solution}
\label{sec2_1}
The temperature $u$ for the surface and the layer beneath is governed by 
\begin{equation}
\label{eq:heat_diffusion}
    \frac{\partial u}{\partial t} = \frac{K}{C \rho } \frac{\partial^2 u}{ \partial z^2},
\end{equation}
with two boundary conditions, 
\begin{align}
    & K \frac{\partial u }{\partial z} \left \vert_{z = 0} = F(t) - e \sigma u^4 \vert_{z=0} \label{eq:BC1}  \right. ,\\
    & K \frac{\partial u}{\partial z} \left \vert_{z \to \infty} = 0, \right. \label{eq:BC2}
\end{align}
and a periodic condition,
\begin{equation}
    u\lvert_{t = 2\pi/\omega} = u\lvert_{t = 0},
\end{equation}
where $t$ is the time, $z$ is the depth of the crater, $\omega$ is the angular velocity, $K$ is the thermal conductivity, $C$ is the specific heat capacity, {{$\rho$ is the bulk density of the asteroid, $e$ is the emissivity, and $\sigma$ is the Stefan-Boltzmann constant}}. In the following, we use the spin angle $\beta = \omega t$ to replace time for simplicity. The effect of the seasonal wave is marginal and is ignored in this work.

While the one-dimensional heat diffusion equation with the boundary condition of radiation has no complete analytical solution, it could be solved by linearizing the fourth power of the temperature, with the assumption that the temperature does not change significantly during a rotational period. For an arbitrary point in the hemispherical crater surface, the location of which is described by the polar angle $\theta$ and the azimuthal angle $\phi$, the solution of the temperature is the real part of the expression
\begin{equation}
    u(\theta,\phi,\beta) = u_0(\theta,\phi) + \sum_{n = 1}^{\infty} u_n(\theta,\phi) {\rm e}^{in\beta},
\end{equation}
with
\begin{align}
    & u_0(\theta,\phi) =  \left( \frac{F_0(\theta,\phi)}{e \sigma } \right)^{1/4}, \label{eq:u0_1} \\
    & u_n(\theta,\phi) =  \frac{F_n(\theta,\phi)}{4e\sigma u_0(\theta,\phi)^3} \frac{{\rm e}^{iJ_n} }{\sqrt{2 \Theta_n^2 + 2 \Theta_n + 1}}, \label{eq:un_1}
\end{align}
where
\begin{align}
    & l_n = \sqrt{\frac{n \omega}{2 M}}, \\
    & \Theta_n = \frac{K l_n}{ 4 e \sigma u_0^3}, \\
    & \tan J_n = -\frac{\Theta_n}{\Theta_n + 1},
\end{align}
{{with $M = K/\rho C$ }}.The function $F_n$ is the $n$th-order of the Fourier series of the absorbed radiation flux $F$, which is expressed as
\begin{equation}
    F(\theta,\phi) =  \sum_{n=0}^\infty F_n(\theta,\phi) {\rm e} ^{i n\beta}
.\end{equation}
We see that the only unknown variable is the absorbed radiation flux $F$. The absorbed radiation on a surface element contains three parts: solar radiation $E(\theta,\phi,\beta),$  radiation from the crater itself $H(\theta,\phi,\beta)$, and the scattering flux from the crater $G(\theta, \phi, \beta),$ which leads us to
\begin{equation}
\label{eq:F}
\begin{aligned}
    F(\theta,\phi) & =  (1-A) (E(\theta,\phi) + H(\theta,\phi) + G(\theta, \phi) )\\
    & =  (1-A) \sum_{n=0}^\infty (E_n(\theta,\phi)+H_n(\theta,\phi) + G_n(\theta, \phi)) {\rm e} ^{i n\beta},
\end{aligned}
\end{equation}
where $E_n(\theta,\phi,\beta)$ and $H_n(\theta,\phi,\beta)$ denote the $n$th-order Fourier modes of $E(\theta,\phi,t)$ and $H(\theta,\phi,t)$, respectively. {{Here the albedo $A$ is assumed to be 0.1 for these three flux components for the sake of simplicity, although the albedo at the thermal-infrared wavelengths is almost zero.}}
Following Equations (\ref{eq:u0_1}), (\ref{eq:un_1}), and (\ref{eq:F}), we obtain
\begin{align}
    & u_0(\theta,\phi) =  \left( \frac{(1-A) (E_0(\theta,\phi)+H_0(\theta,\phi)+G_0(\theta, \phi))}{e \sigma } \right)^{1/4}, \label{eq:u0_2} \\
    & u_n(\theta,\phi) =  \frac{ (1-A)(E_n(\theta,\phi)+H_n(\theta,\phi)+G_n(\theta, \phi))}{4e\sigma u_0(\theta,\phi)^3} \frac{{\rm e}^{iJ_n} }{\sqrt{2 \Theta_n^2 + 2 \Theta_n + 1}}. \label{eq:un_2}
\end{align}
Therefore, to solve the temperature $u(\theta,\phi,\beta)$, we need to obtain the Fourier series of $E(\theta,\phi,\beta)$, $H(\theta,\phi,\beta),$ and $G(\theta,\phi,\beta)$.

\subsubsection{Coordinate systems}
\label{sec:coordinate_system}
We use three coordinate systems to calculate the radiation and the force received by the crater, as shown in Fig~\ref{fig:coordinates}. {{The coordinate system $OXYZ$ is an inertial frame used to calculate the averaged YORP torque over the spin and orbital motion. Based on the axis $OZ$, we define the coordinate system $oxyz$ fixed with the crater to calculate the instant solar flux. Finally, based on the axis $oz$, we define the coordinate system $PABC$ to calculate the self-sheltering effect of the crater. The self-sheltering effect for a point in the crater refers to the non-working moment of the photons reabsorbed by the shelter (i.e. the crater itself), which leads to the effective radiation force of the surface being tilted relative to the surface normal with a modified magnitude. This self-sheltering effect on the point $P$ depends on the geometry of the surrounding shelter. }} We adopt a simple sphere with a radius of $R_1$ to depict the crater shape. 

{{In the coordinate system $OXYZ$, $OZ$ points along the rotation axis and $OXY$ is the equatorial plane. The axis $OY$ is chosen such that the normal vector of the orbit plane lies in the plane $OYZ$. The axis $OX$ follows the right-hand rule.}} There are three crucial vectors determining the CYORP torque expressed in polar coordinates in the coordinate system $OXYZ$: the crater position vector $\vec r_0$ (denoted by $\alpha$ and $\beta$), the crater normal vector $\vec n_0$ (described by  $\alpha$, $\beta$, $\delta,$ and $\Delta$), and the solar position $\vec s$ (described by$\epsilon$ and $\eta$), as shown in the right panel of Fig.~\ref{fig:coordinates}. {{As the polar angle is between $0^\circ$ and $180^\circ$ by definition, we limit $0^\circ< \alpha < 180^\circ$ and $ 0^\circ< \alpha - \Delta < 180^\circ$ in our results.}}

{{The coordinate system $oxyz$ with the origin $o$ located at the sphere centre is fixed with the crater in order to simplify the calculation of the solar flux on the crater. The axis $oz$ points in the opposite direction to the surface normal vector $\vec n_0$, which is also the symmetry axis of the spherical crater. The direction of axis $oy$ is along $\vec e_{OZ} \times \vec e_{oz}$ and $ox$ follows the right-hand rule. In this coordinate system, a crater with a depth of $h$ can be defined as}}
\begin{equation}
\label{eq:crater}
    \mathcal{Z} \coloneqq \{ (x,y,z)  \in \mathbb{R}^3 \lvert  x^2+y^2+z^2 =R_1, z \geq R_1 \cos \theta_0 \},
\end{equation}
{{with $\cos \theta_0 = {(R_1 - h)}/{R_1}$.}}

{{The coordinate system $PABC$ with the origin $P$ at a chosen point on the crater is used to calculate the self-sheltering effect.  The axis $PC$ is along the direction of $Po$, $PA$ points in the direction of $\vec e_{oz} \times \vec e_{PC}$, and $PB$ follows the right-hand rule. In our code, the effective force felt by the point $P$ is calculated first in the coordinate system $PABC$ for simplicity and this is then transformed to the coordinate system $oxyz$ by a rotation matrix.}}

\subsubsection{Solar radiation $E$}
\label{sec2_2}

In this section, we show how we derive the solar radiation received by an arbitrary point $P$ in the crater, whose coordinates in $oxyz$ system are
$\vec r_P = (\sin \theta \cos \phi, \sin \theta \sin \phi, \cos \theta).$
The unit position vector of the Sun in the coordinate system $OXYZ$ is described by
\begin{equation}
    \vec s_{OXYZ} = (\cos \eta, \cos \epsilon \sin \eta, \sin \epsilon \sin \eta),
\end{equation}
{{where $\epsilon$ is the obliquity and $\eta$ is the angle of orbital motion.}}
The transform matrix between the coordinate systems $oxyz$ and $OXYZ$ is set to
\begin{equation}
    \mathcal{R}  = 
\begin{pmatrix}
    \cos \alpha \cos \beta & \cos \alpha \sin \beta & -\sin \alpha\\
    \sin \beta & -\cos \beta & 0\\
    -\cos \beta \sin \alpha & -\sin \alpha \sin \beta & -\cos \alpha
\end{pmatrix}.
\end{equation}
Therefore, the coordinates of the vector $\vec s $ in the coordinate system $oxyz$ is 
\begin{equation}
\begin{aligned}
    & \vec s_{oxyz}  = \\
& \begin{pmatrix}
    -\sin \alpha \sin \epsilon \sin \eta + \cos \alpha \cos \beta \cos \eta + \cos \alpha \sin \beta \cos \epsilon  \sin \eta \\
     \sin \beta \cos \eta - \cos \beta \cos \epsilon \sin \eta\\
    -\sin \alpha \cos \beta \cos \eta  - \sin \alpha \sin \beta \cos \epsilon \sin \eta  - \cos \alpha \sin \epsilon \sin \eta 
\end{pmatrix}.  
\end{aligned}
\end{equation}

On the other hand, we can use the angle $\lambda$ and $\phi'$ to represent $\vec s_{oxyz}$:
\begin{equation}
    \vec s_{oxyz} = (\sin \lambda \cos \phi', \sin \lambda \sin \phi' ,- \cos \lambda),
\end{equation}
such that 
\begin{equation}
    \cos \lambda = \sin \alpha \cos \beta \cos \eta  + \sin \alpha \sin \beta \cos \epsilon \sin \eta  + \cos \alpha \sin \epsilon \sin \eta,
\end{equation}
and 
\begin{equation}
    \tan \phi' = \frac{\sin \alpha \sin \epsilon \sin \eta - \cos \alpha \cos \beta \cos \eta - \cos \alpha \sin \beta \cos \epsilon  \sin \eta}{\sin \beta \cos \eta + \cos \beta \cos \epsilon \sin \eta}.
\end{equation}
The absorbed radiation flux is 
\begin{equation}
\label{eq:E0}
\begin{aligned}
    E(\theta,\phi) & = (1-A) \Phi  H(\cos \lambda) H(w)  (-\vec r_P \cdot \vec s) \\
    & = (1-A) \Phi  H(\cos \lambda) H(w) \cdot  [m_1 \cos(\beta - \beta_1) + m_2]
\end{aligned}
\end{equation}
where
\begin{equation}
    w = \cos 2\lambda \cos \theta + \sin \theta_0 - \sin 2\lambda \sin \theta \cos (\phi - \phi'),
\end{equation}
\begin{equation}
\begin{aligned}
    &\tan \beta_1 = \\
    & \frac{ (\sin \alpha \cos \theta - \cos \alpha \sin \theta \cos \phi) \cos \epsilon \sin \eta - \sin \theta \sin \phi \cos \eta}{(\sin \alpha \cos \theta - \cos \alpha \sin \theta \cos \phi) \cos \eta + \cos \epsilon \sin \eta \sin \theta \sin \phi},
\end{aligned}
\end{equation}
\begin{equation}
\begin{aligned}
    & m_1 = (\cos^2\eta + \cos^2\epsilon \sin^2\eta)^{1/2} \\
    & \cdot (\cos^2\theta \sin^2\alpha - \frac{\cos\phi \sin2\alpha \sin 2\theta}{2} \\
     & + (\cos^2\alpha \cos^2\phi + \sin^2\phi) \sin^2\theta)^{1/2},
\end{aligned}
\end{equation}
\begin{equation}
    m_2 = \cos \alpha \cos \theta \sin \epsilon \sin \eta + \cos \phi \sin \alpha \sin \epsilon \sin \eta \sin \theta.
\end{equation}
Here, $H$ is the Heaviside function defined by
\begin{equation}
    H(x) \coloneqq 
    \left \{
    \begin{aligned}
    & 1, x>0 \\
    & 0, \rm else.
    \end{aligned}
    \right.
\end{equation}
Using the substitution $\beta' = \beta - \beta_1$, the absorbed radiation flux has the form
\begin{equation}
\label{eq:E1}
    E(\theta,\phi) = 
    \left \{
    \begin{aligned}
    & (1-A)\Phi (m_1 \cos\beta' + m_2), \beta'_{\min} < \beta < \beta'_{\max} \\
    & 0, \rm else.
    \end{aligned}
    \right.    
\end{equation}
Expanding Equation (\ref{eq:E1}) in Fourier series, we obtain
\begin{equation}
    E(\theta,\phi) = E_0 + \frac{(1-A)\Phi}{\pi} \sum_{n=1}^{\infty} \left [ C_n \cos(n \beta') + S_n \sin (n \beta') \right],
\end{equation}
with 
\begin{equation}
    E_0 = \frac{(1-A)\Phi}{2\pi} (m_1 \sin \beta + m_2 \beta) \left|_{\beta'_{\min}}^{\beta'_{\max}}  \right. ,
\end{equation}
\begin{equation}
\begin{aligned}
    C_n & =   \int_{\beta'}^{\beta_{\min}'} (m_1 \cos\beta_{\max}' + m_2) \cos (n \beta'){\rm d} \beta'  \\
    & = 
    \left \{
    \begin{aligned}
    & \left(\frac{m_1 \sin 2\beta' + 2 m_1 \beta'}{4} + m_2 \sin \beta \right)  \left |_{\beta_{\min}'}^{\beta_{\max}'} \right. , n = 1\\
    & \left ( \frac{m_1 n \cos \beta' \sin n \beta' - m_1 \cos n\beta'  \sin \beta'}{n^2 -1} + \frac{m_2 \sin (n\beta')}{n}  \right)     \left |_{\beta_{\min}'}^{\beta_{\max}'} \right.,\\
    &  \qquad \qquad n>1,
    \end{aligned}
    \right.    
\end{aligned}
\end{equation}
and
\begin{equation}
\begin{aligned}
    S_n & =   \int_{\beta'}^{\beta_{\min}'} (m_1 \cos \beta_{\max}' + m_2) \sin (n \beta'){\rm d} \beta'  \\
    & = 
    \left \{
    \begin{aligned}
    &  - \left (\frac{m_1 \cos \beta'^2}{2} + m_2 \cos \beta' \right)  \left |_{\beta_{\min}'}^{\beta_{\max}'} \right. , n = 1\\
    &  \left ( \frac{m_1 n \cos \beta' \cos n \beta' + m_1 \sin \beta' \sin n \beta'}{1 - n^2} - \frac{m_2 \cos n\beta'}{n}  \right)     \left |_{\beta_{\min}'}^{\beta_{\max}'} \right.,\\
    &  \qquad \qquad  n>1.
    \end{aligned}
    \right.    
\end{aligned}
\end{equation}
The $n$-th coefficient of the Fourier series of $E(\theta, \phi)$ is 
\begin{equation}
    E_n = \sqrt{C_n^2 + S_n^2} \cdot {\rm e}^{i \Phi_n}
,\end{equation}
with 
\begin{equation}
    \tan \Phi_n = -\frac{S_n}{C_n}.
\end{equation}

\subsubsection{Self-heating effect}
\label{sec2_3}

Due to the concave structure of the crater, the surface element in the crater also receives photons emitted by the crater itself, which is a process referred to as  `self-heating'. In this section, we derive the self-radiation $H$ and the self-scattering $G$ as a function of the position in the crater.

Let us consider a surface element d$S_1$ receiving the radiation from another surface element d$S_2$. In the reference frame $oxyz$, the positions of d$S_1$ and d$S_2$ are 
\begin{align}
    \vec r_1 &= R_1(\sin \theta_1 \cos \phi_1, \sin \theta_1 \sin \phi_1, \cos \theta_1), \label{eq:r_1} \\
    \vec r_2 &= R_1(\sin \theta_2 \cos \phi_2, \sin \theta_2 \sin \phi_2, \cos \theta_2) .
\end{align}
The displacement from d$S_1$ to d$S_2$ is 
\begin{equation}
    \vec r_{1,2} = \vec r_2 - \vec r_1.
\end{equation}
The incident angle $\zeta_1$ and the emission angle $\zeta_2$ are defined as
\begin{align}
    &\cos \zeta_1 = -\vec r_1 \cdot \vec r_{1,2}/R_1 r_{1,2}, \\
    &\cos \zeta_2 = \vec r_2 \cdot \vec r_{1,2}/R_1 r_{1,2}, \label{eq:zeta_2}
\end{align}
respectively. 
The radiation flux at the location of d$S_1$ produced by the thermal radiation of d$S_2$ is 
\begin{equation}
\label{eq:H12}
    H_{1,2} = \frac{ e\sigma u_2^4}{\pi} \frac{\cos \zeta_1 \cos \zeta_2}{\vec r_{1,2}^2} dS_2.
\end{equation}
Substituting Equations (\ref{eq:r_1})-(\ref{eq:zeta_2}) into Equation (\ref{eq:H12}), we obtain
\begin{equation}
\begin{aligned}
    H_{1,2} &= \frac{ e \sigma u_2^4}{4\pi R_1^2}dS_2.
\end{aligned}
\end{equation}
For an arbitrary point, the radiation flux caused by the whole crater is
\begin{equation}
\begin{aligned}
    H(\theta, \phi) &= \int_{\mathcal{Z}} \frac{ e\sigma u(\theta',\phi')^4}{4\pi}  \sin \theta' d\theta' d\phi',
\end{aligned}
\end{equation}
where $\mathcal{Z}$ is the crater surface, and is defined as
\begin{equation}
    \mathcal{Z} \coloneqq \{ (x,y,z)  \in \mathbb{R}^3 \lvert  r =R_1, \theta \in (0,\pi/2 - \theta_0), \phi \in (0,2\pi) \}.
\end{equation}
Similarly, we obtain the self-scattering term:
\begin{equation}
    G(\theta, \phi) = \int_{\mathcal{Z}} \frac{A E(\theta',\phi')}{4\pi}  \sin \theta' d\theta' d\phi'.
\end{equation}
Therefore, both $H(\theta, \phi)$ and $G(\theta, \phi)$ can be expressed in terms of $u(\theta, \phi)$ and $E(\theta, \phi)$. As $E(\theta, \phi)$ is derived in Sect. \ref{sec2_2}, the only unknown variable is the temperature $u(\theta, \phi)$, the solution for which is discussed in the following section.

\subsection{Solution for temperature}
\label{sec2_4}

We obtained the Fourier series of the solar radiation flux (Sect. \ref{sec2_2}) and the radiation flux produced by the crater (Sect. \ref{sec2_3}), which allows us to return to Equation (\ref{eq:F}) to solve the temperature distribution $u(\theta, \phi)$. We note that the self-radiation term $H$ contains the unknown temperature distribution that is to be solved. 

The Fourier coefficients of the temperature of the crater follow
\begin{equation}
\label{eq:sigma_u}
\begin{aligned}
    & \sigma u_0^4(\theta,\phi) =  (1-A)(E_0(\theta,\phi) + G_0(\theta, \phi) + H_0(\theta, \phi)) \\
    &  \sigma u_n(\theta,\phi) u_0^3(\theta,\phi) \frac{2{\sqrt{2 \Theta_n^2 + 2 \Theta_n + 1}}}{ (1-A){\rm e}^{iJ_n}} = E_n(\theta,\phi) + G_n(\theta, \phi) + H_n(\theta, \phi),
\end{aligned}
\end{equation}
with
\begin{equation}
\begin{aligned}
    & G_n(\theta, \phi) = \int_{\mathcal{Z}} \frac{A E_n(\theta',\phi')}{4\pi} \sin \theta' d\theta' d\phi', \\
    & H_0(\theta, \phi) = \int_{\mathcal{Z}} \frac{ e\sigma u_0(\theta',\phi') ^4} {4\pi}\sin \theta' d\theta' d\phi', \\
    & H_n(\theta, \phi) = \int_{\mathcal{Z}} \frac{ e \sigma u_0(\theta',\phi') ^3 u_n(\theta',\phi')}{\pi}\sin \theta' d\theta' d\phi'.
\end{aligned}
\end{equation}
Here, terms $G$ and $H$ are the scattering flux and self-radiation flux, respectively. 

\subsubsection{Solution of a general form}
Equations (\ref{eq:sigma_u}) have a general form:
\begin{equation}
\label{eq:f1}
    f(\theta, \phi) = g(\theta, \phi) + C \int_{\mathcal{Z}} f(\theta', \phi') \sin \theta' d \theta' d\phi'.
\end{equation}
By setting 
\begin{equation}
\label{eq:k1}
    k = \int_{\mathcal{Z}} f(\theta', \phi') \sin \theta' d \theta' d\phi',
\end{equation}
we have
\begin{equation}
\label{eq:f2}
    f(\theta, \phi) = g(\theta, \phi) + C \cdot k.
\end{equation}
Substituting Equation (\ref{eq:f2}) into Equation (\ref{eq:k1}), we obtain
\begin{equation}
\label{eq:k2}
    k = \int_{\mathcal{Z}} (g(\theta', \phi') + C \cdot k) \sin \theta' d \theta' d\phi'.
\end{equation}
Rearranging Equation (\ref{eq:k2}), we find $k$ is 
\begin{equation}
    k = \frac{\int_{\mathcal{Z}} g(\theta', \phi') \sin \theta' d \theta' d\phi' }{ 1 - C \int_{\mathcal{Z}}  \sin \theta' d \theta' d\phi'},
\end{equation}
with which $f(\theta, \phi)$ is solved out by Equation (\ref{eq:f2}).

\subsubsection{Solution for temperature}
In the case of $u_0$, 
\begin{align}
    & f(\theta, \phi) = e\sigma u_0^4 /(1-A), \\
    & g(\theta, \phi) = E_0(\theta, \phi) + G_0(\theta, \phi), \\
    & C = \frac{1}{4 \pi},
\end{align}
and in the case of $u_n$,
\begin{align}
    & f(\theta, \phi) = e\sigma u_0^3 u_n \frac{2{\sqrt{2 \Theta_n^2 + 2 \Theta_n + 1}}}{(1-A){\rm e}^{iJ_n}}, \\
    & g(\theta, \phi) = E_n(\theta, \phi) + G_n(\theta, \phi), \\
    & C = \frac{ {\rm e}^{i J_n}}{2 \pi \sqrt{2 \Theta_n^2 + 2 \Theta_n + 1}}.
\end{align}

\subsection{The CYORP torque}
The CYORP  torque is defined as the YORP torque difference between the crater and the flat ground:
\begin{equation}
\label{eq:T_CYORP}
    T_{\rm CYORP} = T_{\rm crater} - T_{\rm ground}.
\end{equation}

\subsubsection{The YORP torque of the crater}
The average radiative torque produced by the crater should be calculated in the inertial frame $OXYZ$:
\begin{equation}
    \vec T_{\rm crater} =\vec r_{0, OXYZ} \times \vec f_{OXYZ}.
\end{equation}
Here, $\vec f_{OXYZ}$ is the radiative force, and
\begin{equation}
    \vec f_{OXYZ} =  \int_{\mathcal{W}}  \frac{e \sigma T(\theta, \phi)^4 }{ c} \vec n_f(\theta, \phi) dS,
\end{equation}
where $\vec n_f$ is the corrected force direction vector for each surface element. If there is no shelter around the surface element, $\vec n_f$ is equal to the surface unit normal vector $\vec n_0$. However, the surface element in a crater has a sky sheltered by other elements, resulting in the reabsorption of the emitted photons along the direction of the shelter.

\begin{figure*}
    \centering
    \includegraphics[width = \textwidth]{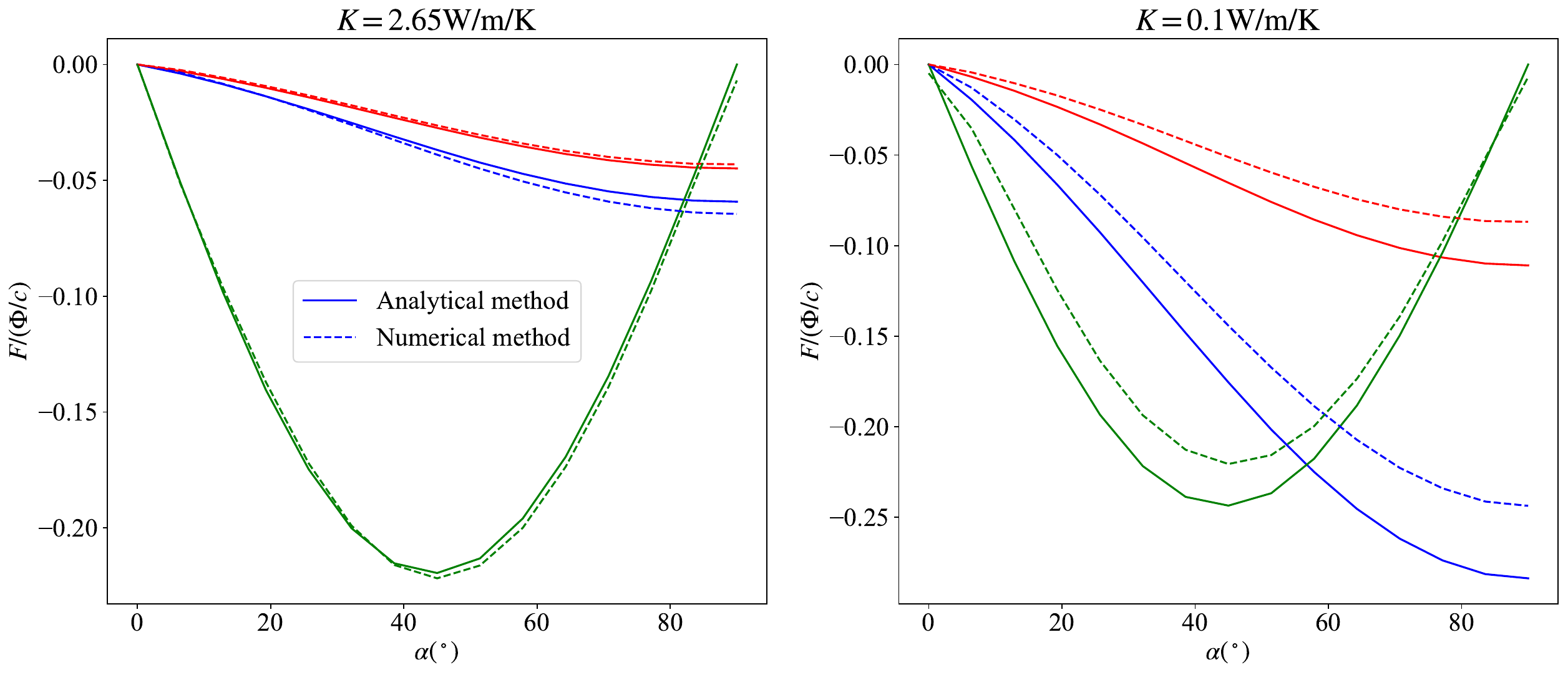}
    \caption{Radiative force of the total crater averaged over a rotational period {{(eight hours by default)}} obtained from the analytical method (solid lines) and the numerical method (dashed lines), as a function of the crater colatitude $\alpha$ for $K = 2.65 \rm W/m/K$ (left panel) and $K = 0.1 \rm W/m/K$ (right panel). The $x$, $y$, and $z$ components of the radiative force are shown in blue, red, and green, respectively. {{The distance of the crater here is 1 au from the Sun.}}}
    \label{fig:comparison}
\end{figure*}

\begin{figure}
    \centering
    \includegraphics[width = 0.5\textwidth]{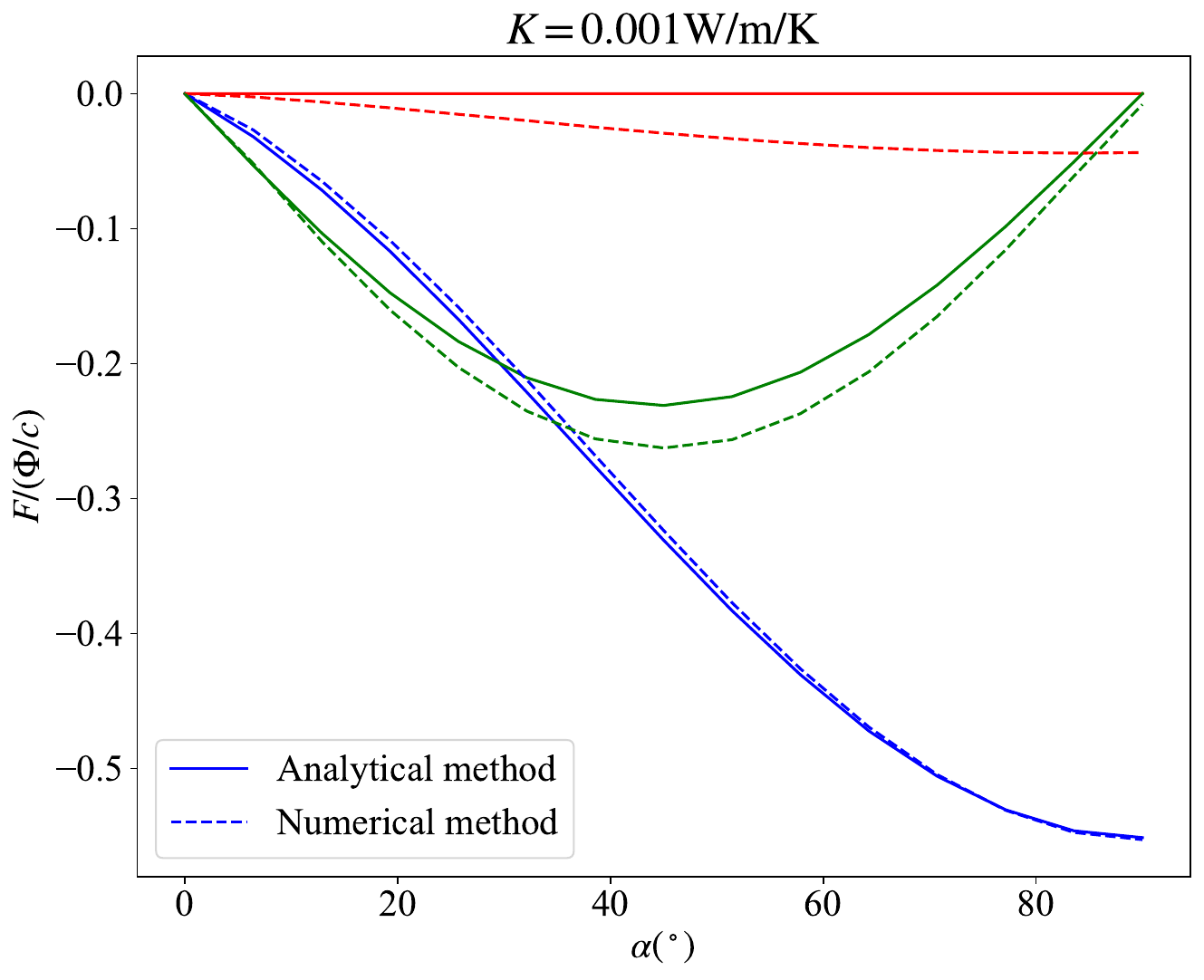}
    \caption{Same as Fig.~\ref{fig:comparison} but for $K = 0 \rm W/m/K$ in the analytical model and $K = 0.001 \rm W/m/K$ in the numerical model.}
    \label{fig:comparison2}
\end{figure}

For the surface element dS($\theta$, $\phi$), the radiative force is 
\begin{equation}
    \vec{f} = -\int_{\mathcal{H}} \frac{\Phi}{\pi c} \cos \theta'
    \left (
    \begin{aligned}
    &\sin \theta' \cos \phi' \\
    &\sin \theta' \sin \phi' \\
    &\cos \theta'
    \end{aligned}
    \right )  \sin \theta'  {\rm d}\theta' {\rm d}\phi'.
\end{equation}
Without sheltering (e.g. for convex asteroids), $\mathcal{H}$ is replaced by the hemisphere (i.e. $\theta \in (0, \pi/2)$, $\phi \in (0, 2 \pi)$). In this case, the force is reduced to $2\Phi \vec n_0/3c$.

\subsubsection{The YORP torque of the flat portion of the surface}
The absorbed radiation flux for a flat ground with the normal vector $\vec n_0$ is 
\begin{equation}
\begin{aligned}
    E_{\rm ground} & = (1-A) \Phi H(\cos \lambda) \cos \lambda \\
    & = \left \{
    \begin{aligned}
    & (1-A)\Phi (m_1 \cos\beta' + m_2), \beta'_{\min} < \beta' < \beta'_{\max} \\
    & 0, \rm else.
    \end{aligned}
    \right . 
\end{aligned}
\end{equation}
Here, $\beta' = \beta - \beta_1$, with
\begin{equation}
    \tan \beta_1 = \frac{ \sin \alpha \cos \epsilon \sin \eta}{\sin \alpha \cos \eta},
\end{equation}
\begin{align}
    & m_1 = ( \sin^2 \alpha (\cos^2 \eta + \cos^2 \epsilon \sin^2 \eta   )  )^{1/2}, \\
    & m_2 = \cos \alpha \sin \epsilon \sin \eta,
\end{align}
and $\beta_{\rm min}$ and $\beta_{\rm max}$ are the negative and positive values of ${\rm arccos} (- m_2/m_1)$, respectively.

\section{Numerical model for examination}
\label{sec3}

In the above analytical method, we adopted the assumption of a “small” temperature variation during a rotation period, which allows us to linearize the biquadrate of the temperature (i.e. $u^4 \sim u_0^3 \sum u_n {\rm e}^{\beta n {\rm i}} $). This assumption is equivalent to a high value of the thermal parameter, which is defined as
\begin{equation}
\Gamma = \frac{ \sqrt{C \rho \omega \kappa}}{(\epsilon \sigma)^{1/4} (1-A)^{3/4} \Phi^{3/4}}.
\end{equation}

In the case of a low thermal parameter, the analytical model should be used with caution. To examine the appropriate range of the thermal inertia for which our analytical model is valid, we built a one-dimensional thermophysical numerical model to perform cross-validation. 

\subsection{Numerical model}

We used a finite difference numerical method to solve Equation \ref{eq:heat_diffusion} with the second-order Crank-Nicholson scheme:
\begin{equation}
\label{eq:u_ijk}
\begin{aligned}
     c_{\rm n} u_{i,j+1,k+1} - (2 c_{\rm n} + 1) u_{i,j,k+1} + c_{\rm n} u_{i,j-1,k+1} \\
     = -c_{\rm n} u_{i,j+1,k} + (2c_{\rm n}-1) u_{i,j,k} - c_{\rm n} u_{i,j-1,k}.   
\end{aligned}
\end{equation}
Here, $u_{i,j,k}$ represents the temperature at the depth of $(j-1) \delta z$ below the $i$-th facet at the $k$-th time step, where $i, j,$ and $k$ are integrals starting from 1 to $i_{\rm max}, j_{\rm max},$ and $k_{\rm max}$. The coefficient $c_{\rm n}$ is
\begin{equation}
    c_{\rm n} = \frac{a \delta t}{2(\delta x)^2}.
\end{equation}
The value of $c_{\rm n}$ should be smaller than 0.5 for the convergence of the iteration.

The surface temperature is determined by the first boundary condition (Equation \ref{eq:BC1}):
\begin{equation}
\label{eq:BC1_numerical}
    (1-A)(E_{i,k} + H_{i,k} + S_{i,k}) - \sigma u_{i,1,k+1}^4 =  K \frac{u_{i,1,k+1} - u_{i,2,k}}{\delta z},
\end{equation}
which can be solved by a Newtonian-Raphson iterative method. The solar flux $E_{i,k}$ on the $i$-th facet at the $k$-th time step is 
\begin{equation}
    E_{i,k} =
    \left \{
    \begin{aligned}
    & \Phi_0 \vec n_i \cdot \vec s_k H(\vec n_i \cdot \vec s_k), {\rm unsheltered} \\
    & 0, {\rm sheltered,}  
    \end{aligned}
    \right.    
\end{equation}
where $\Phi_0$ is 1364 $\rm W/m^2$ at the distance of 1 au to the Sun. Here, $\vec n_i$ is the normal vector of the $i$-th facet and $\vec s_k$ is the unit solar position vector at the $k$-th time step. Whether or not the facet is sheltered is judged according to the projections of other facets on the plane of the $i$-th facet along the solar position vector $\vec s_k$. The self-radiation term $H_{i,k}$ is the sum of radiation from other facets:
\begin{equation}
    H_{i,k}  = \sum_{i' \neq i} e \sigma u_{i',1,k}^4 \frac{-(\vec n_i \cdot \vec r_{i,i'}) (\vec n_{i'} \cdot \vec r_{i,i'}) }{\pi r_{i,i'}^2} S_{i'},
\end{equation}
where $\vec r_{i,i'}$ is the vector from the centre of the $i$-th facet to the centre of $i'$-th facet, and $S_{i'}$ is the area of the $i$-th facet. The scattering term $S_{i,k}$ is given by
\begin{equation}
    S_{i,k} = \sum_{i' \neq i} A E_{i'} \frac{-(\vec n_i \cdot \vec r_{i,i'}) (\vec n_{i'} \cdot \vec r_{i,i'}) }{\pi r_{i,i'}^2} S_{i'}.
\end{equation}

The second boundary condition (Eq. \ref{eq:BC2}) translates into 
\begin{equation}
\label{eq:BC2_numerical}
    u_{i,j_{\rm max},k+1} = u_{i,j_{\rm max}-1,k+1}.
\end{equation}
Combining Equations (\ref{eq:u_ijk}), (\ref{eq:BC1_numerical}), and (\ref{eq:BC2_numerical}), we can obtain the solution for the temperature in the next time step $u_{i,j,k+1}$ based on the temperature in the current time step $u_{i,j,k}$. We adopted an initial temperature of $u_{i,j,0} = 280~\rm K$. The maximum depth is set to be a few thermal skin depths $\sqrt{K/C\rho\omega}$ and the total number of layers is set as $j_{\rm max} = 50$. In order to make sure that the surface temperature is in an equilibrium state, we set the total time as $k_{\rm max}\delta t \sim 20$ spin periods. We divided the considered crater into about $ 1000$ facets and solve the temperature for each facet using the above method. 

\subsection{Comparison with the analytical model}
The thermal parameters of asteroids can vary widely depending on their composition and structure. For example, the thermal conductivity of a porous material is much lower than that of a dense metal. The thermal conductivity of stony asteroids, which are made mostly of silicates, can range from about 0.1 W/mK to 1 W/mK, while the thermal conductivity of metallic asteroids is generally much higher, in the range of 20-50 W/mK. Asteroids that are composed of a mixture of rock and metal will have thermal conductivity values between those of pure rock and pure metal.

Here, we test three typical types of asteroid materials: regolith, solid basalt, and metal, whose properties are shown in Table \ref{tab1}. We calculate the radiative force of the total crater averaged over a rotational period {{(eight hours by default)}}, as a function of the colatitude of the crater. {{The craters in the test are placed at 1 au from the Sun.}} For simplicity, we set the obliquity to $\epsilon = 0$. The results computed from the analytical method and the numerical method are shown in Fig.~\ref{fig:comparison}. We can see that the analytical result is consistent with the numerical result to a high degree, while a large discrepancy shows up when the thermal conductivity decreases to 0.1~W/m/K. This coincides with our prerequisite of the application of our analytical method, which is that the temperature variation should be small. Therefore, our method is appropriate for solid basalt and metal materials. 

Regarding regolith material, with a thermal conductivity of as low as 0.001~W/m/K, we test the model described in \citet{Zhou2022}, where zero thermal conductivity is assumed. The result is shown in Fig.~\ref{fig:comparison2}, which indicates that this latter model works well for regolith materials. Therefore, for three materials representing asteroid surfaces, our two methods, namely the one in the present work and that described in \citet{Zhou2022}, behave well in modelling the YORP effect.

\begin{figure}
    \centering
    \includegraphics[width = 0.5\textwidth]{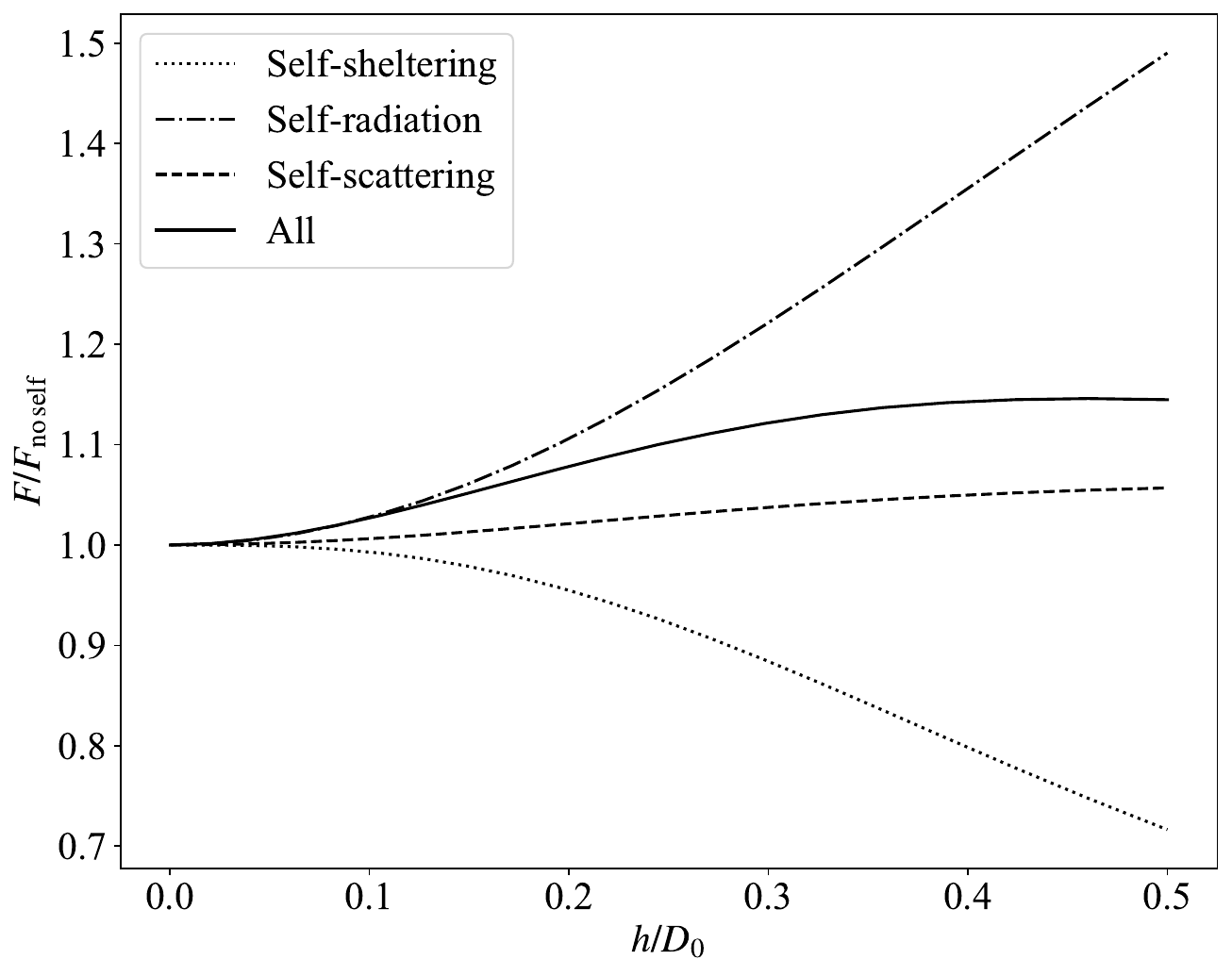}
    \caption{Ratio of the radiative force including self-modification effects to that without self-modification effects, as a function of crater depth-to-diameter ratio and accounting for different self-modification effects. }
    \label{fig:self_modification}
\end{figure}

\begin{table}[]
\caption{Thermal parameter for three typical materials on asteroids, taken from \citet{Farinella1998}.}
    \centering
    \begin{tabular}{c c c c}
         \hline
         \noalign{\smallskip}
         & $\rho~(\rm kg/m^3)$ & $K~\rm (W/m/K)$ & $C~\rm (J/kg/K)$ \\
         \noalign{\smallskip}
         \hline
         \noalign{\smallskip}
         Regolith & 1500 & 0.0015 & 680 \\
         Solid basalt & 3500 & 2.65 & 680 \\
         Solid iron & 8000 & 40 & 500 \\
         \noalign{\smallskip}
         \hline
    \end{tabular}
    \label{tab1}
\end{table}

\section{Discussion on self-modification effects}
\label{sec4}

For a concave structure, there are three self-modification effects: the self-sheltering, self-radiation, and self-scattering effect. The first one refers to the radiative force modification on the surface element due to the re-absorption of photons by the crater. The second and third ones refer to the temperature increase due to the emitted photons and scattered photons from the crater itself. {{Previous research shows that these self-modification effects could be important for the YORP torque of the crater \citep{Statler2009, Rozitis2012, Rozitis2013a}, but a quantitative description is still lacking. {For example, it is not clear how deep the crater needs to be so that these self-modification effects can no longer be ignored. This is crucial to the validity of the commonly made assumption that the asteroid can be considered as a moderately convex shape with the craters being overlooked when evaluating the YORP torque.} An essential metric in this respect is the depth-to-diameter ratio of the crater. We explore the radiative force (leading to the YORP torque) of a crater based on its depth-to-diameter ratio under varying conditions—one for each self-modification effect. Figure~\ref{fig:self_modification} shows our findings, situating the craters at the equator ($\alpha = \pi/2$) with a  spin axis of zero-degree obliquity. }}

{{Our analysis reveals that both self-radiation and self-scattering amplify the force, raising the temperature by as much as 50\% when the crater mirrors a hemisphere. In contrast, self-sheltering diminishes the force by up to 30\%. Notably, the impact of self-scattering remains negligible for typical asteroid surface albedos ranging between 0.1 and 0.2. When integrating all self-modification effects, the radiative force increases by 15\%.}} 

{{Our result also shows that when the depth-to-diameter ratio $h/D_0 < 0.05$, the force increase is less than 1\%. Therefore, for those shallow concave structures with $h/D_0 < 0.05$, no self-modification effects are needed in the YORP model for it to remain accurate. These can then be efficiently approximated as flat surfaces.
}}

\section{Analysis of the CYORP torque}
\label{sec5}

As shown in a previous work \citep{Zhou2022}, the CYORP torque depends on many factors, including the depth-to-diameter ratio and location (or the colatitude $\alpha$ for example) of the crater as well as the obliquity and thermal parameter of the asteroid. In this section, we discuss the dependence of the CYORP torque on these factors. {{In the following, except for in Sect.~\ref{sec:depth_to_diameter_ratio},}} we assume the depth-to-diameter ratio to be $\sim 0.16$.

\subsection{Depth-to-diameter ratio}
\label{sec:depth_to_diameter_ratio}
\begin{figure*}
    \centering
    \includegraphics[width=\textwidth]{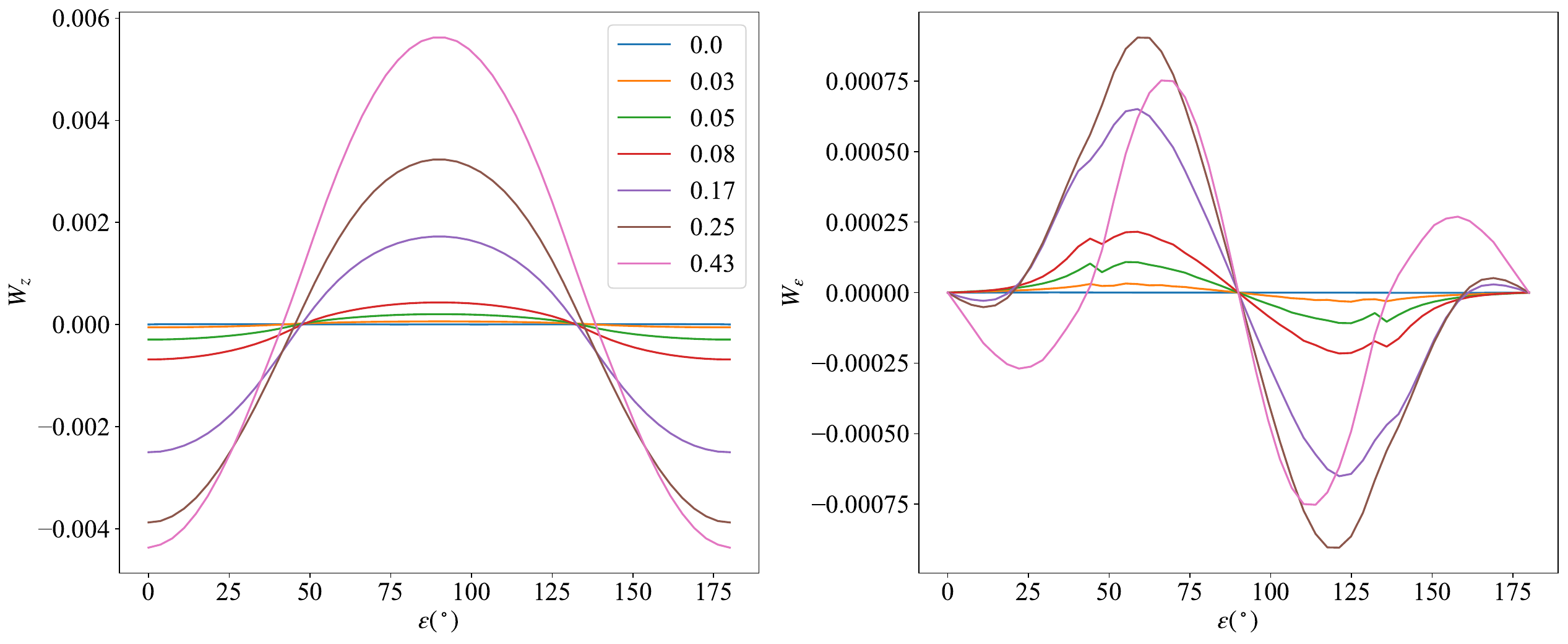}
    \caption{Spin component (left panel) and the obliquity component (right panel) of the CYORP coefficient as a function of the asteroid obliquity $\epsilon$, accounting for different depth-to-diameter ratios, which are denoted by different colours.}
    \label{fig:depth_ratio}
\end{figure*}

Asteroid craters exhibit a range of distinct features in size and shape, with diameters ranging from a few centimetres to several kilometres for large asteroids. These craters generally display bowl-shaped structures, containing central peaks and terraced walls when produced in the gravity regime. To simplify the modelling, a semi-sphere approximation is often used to represent the shape of craters. 

According to the definition of
CYORP torque, if the depth-to-diameter ratio reaches zero, the CYORP torque is zero  (Eq. \ref{eq:T_CYORP}). Figure~\ref{fig:depth_ratio} shows the CYORP torques generated by craters with various depth-to-diameter ratios. The parameters $\delta = \Delta = \pi/6$, $K = 2.65~\rm W/m/K$, and $\alpha = \pi/2$ are used. Higher depth-to-diameter ratios correspond to larger spin components of the CYORP torque. A crater with $h/D_0 < 0.05$ produces an insignificant CYORP torque, which may be disregarded. Furthermore, the depth-to-diameter ratio also impacts the obliquity component, influencing both the torque magnitude and shape of the torque curve. For instance, when the depth-to-diameter ratio is low, the asymptotic obliquity is $90^\circ$, while for higher depth-to-diameter ratios, new asymptotic obliquities arise around $0^\circ$ and $180^\circ$.

\subsection{Crater latitude $\alpha$ and asteroid obliquity $\epsilon$}
\label{sec5_2}

\begin{figure*}
    \centering
    \includegraphics[width = \textwidth]{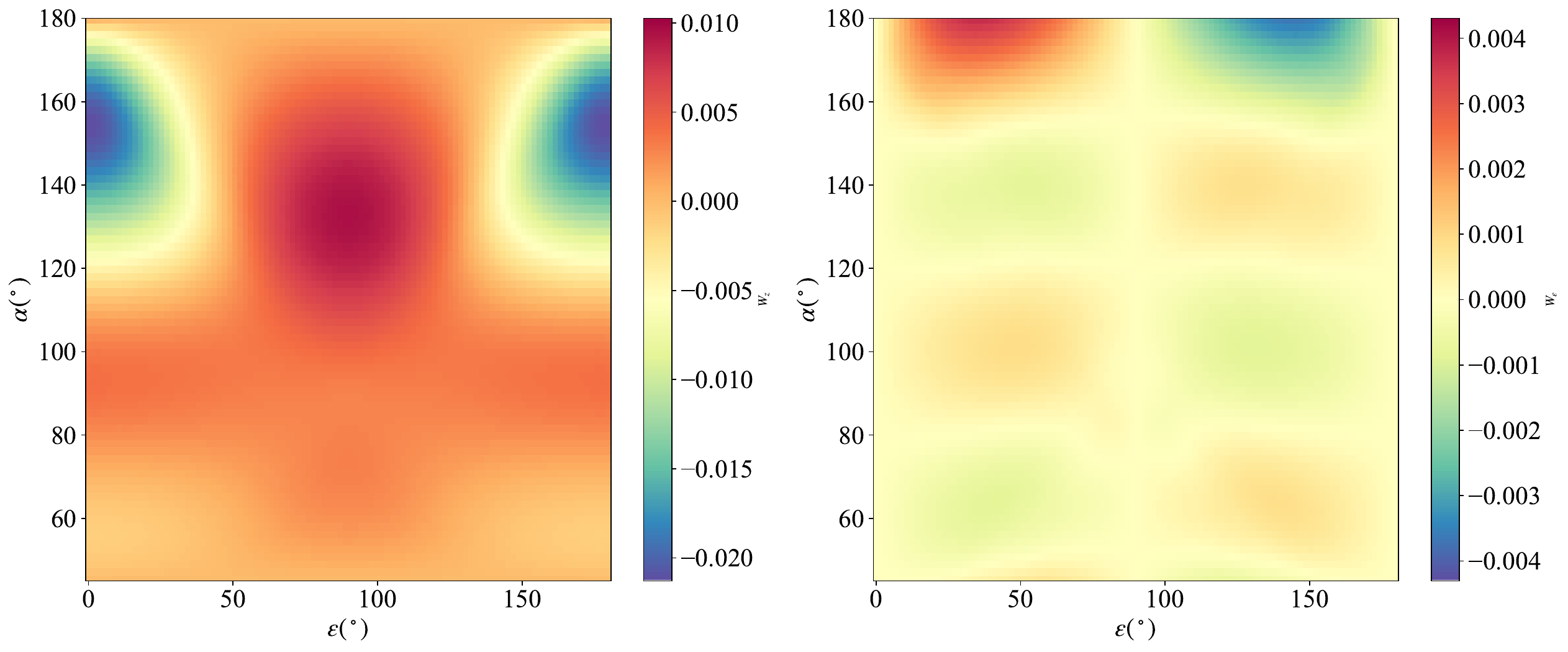}
    \caption{Spin component (left panel) and the obliquity component (right panel) of the CYORP coefficient as a function of the asteroid obliquity $\epsilon$ and the crater colatitude $\alpha (> \Delta)$, for $K = 2.65\rm W/m/K$. Here, $d/D_0 = 0.16$ and $\delta = \Delta = \pi/4$.}
    \label{fig:alpha_epsilon}
\end{figure*}

\begin{figure*}
    \centering
    \includegraphics[width=\textwidth]{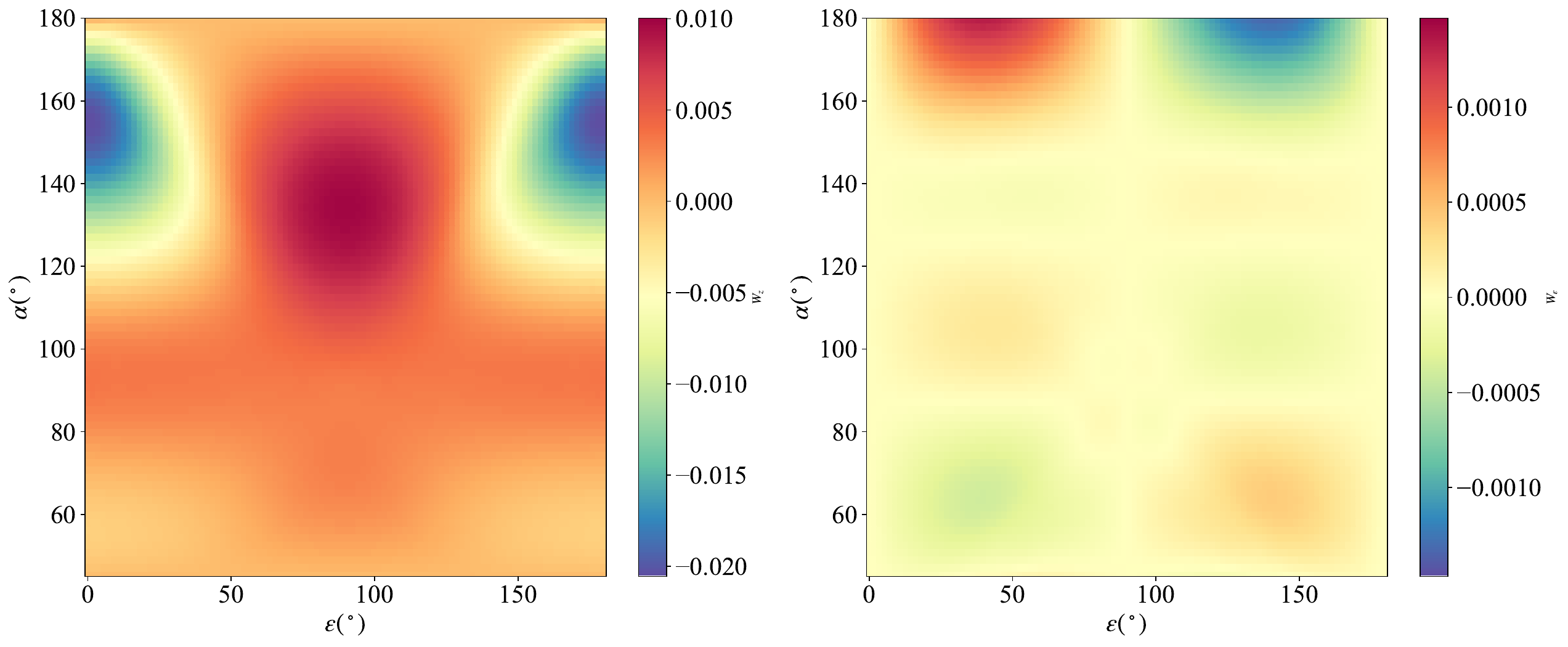}
    \caption{Same as Fig. \ref{fig:alpha_epsilon} but for $K = 40\rm W/m/K$.}
    \label{fig:thermal_inertia}
\end{figure*}

Figure \ref{fig:alpha_epsilon} displays the CYORP torque components as a function of the crater latitude and asteroid obliquity. The values of the parameters $\delta$ and $\Delta$ are set to a representative value of $\pi/4$. The spin component $W_{z}$ exhibits symmetry about the axis $\epsilon = 90^\circ$, while the obliquity component $W_\epsilon$ is anti-symmetric. The minimum and maximum values of the torque occur when the obliquity is $0\circ$ or $90^\circ$, with the absolute value of these extrema reaching up to 0.02. The coefficient of the obliquity component of the CYORP torque is considerably smaller, with a maximum value of 0.004.

For comparison, the typical value of the normal YORP spin coefficient is 0.005 for type \uppercase\expandafter{\romannumeral1}/\uppercase\expandafter{\romannumeral2} and $<0.001$ for type \uppercase\expandafter{\romannumeral3}/\uppercase\expandafter{\romannumeral4} asteroids \footnote{See the definition by \citep{Vokrouhlicky02}}. The ratio of the CYORP torque to the normal YORP torque scales as 
\begin{equation}
    \frac{T_{\rm CYORP}}{T_{\rm NYORP}} = \frac{W_{\rm CYORP}S_{\rm crater}}{W_{\rm YORP}S_{\rm asteroid}}.
\end{equation}
Setting the ratio to 1, we find that the total area of concave structures needs to be as large as 1/4 and 1/20 of the asteroid surface area for type \uppercase\expandafter{\romannumeral1}/\uppercase\expandafter{\romannumeral2} and type \uppercase\expandafter{\romannumeral3}/\uppercase\expandafter{\romannumeral4} asteroids, respectively.

\subsection{Thermal parameter}

When the asteroid rotates quickly and has high heat conductivity, a higher value of the thermal parameter arises, resulting in a smaller variation in temperature. To explore the role of the thermal parameter, we employ the same parameters as in Sect.~\ref{sec5_2}, but with $K=40~\rm W/m/K$ for metal materials, and the resulting CYORP torques are depicted in Fig.~\ref{fig:thermal_inertia}. The comparison with Fig.~\ref{fig:alpha_epsilon} reveals that the spin component remains relatively unchanged, while the obliquity component displays significant variation. This observation aligns with the prior assertion that the thermal parameter mainly impacts the obliquity component \citep{Vokrouhlicky02}. In the regime of high thermal conductivity, the obliquity component diminishes as the thermal conductivity increases.

\begin{figure*}
    \centering
    \includegraphics[width = \textwidth]{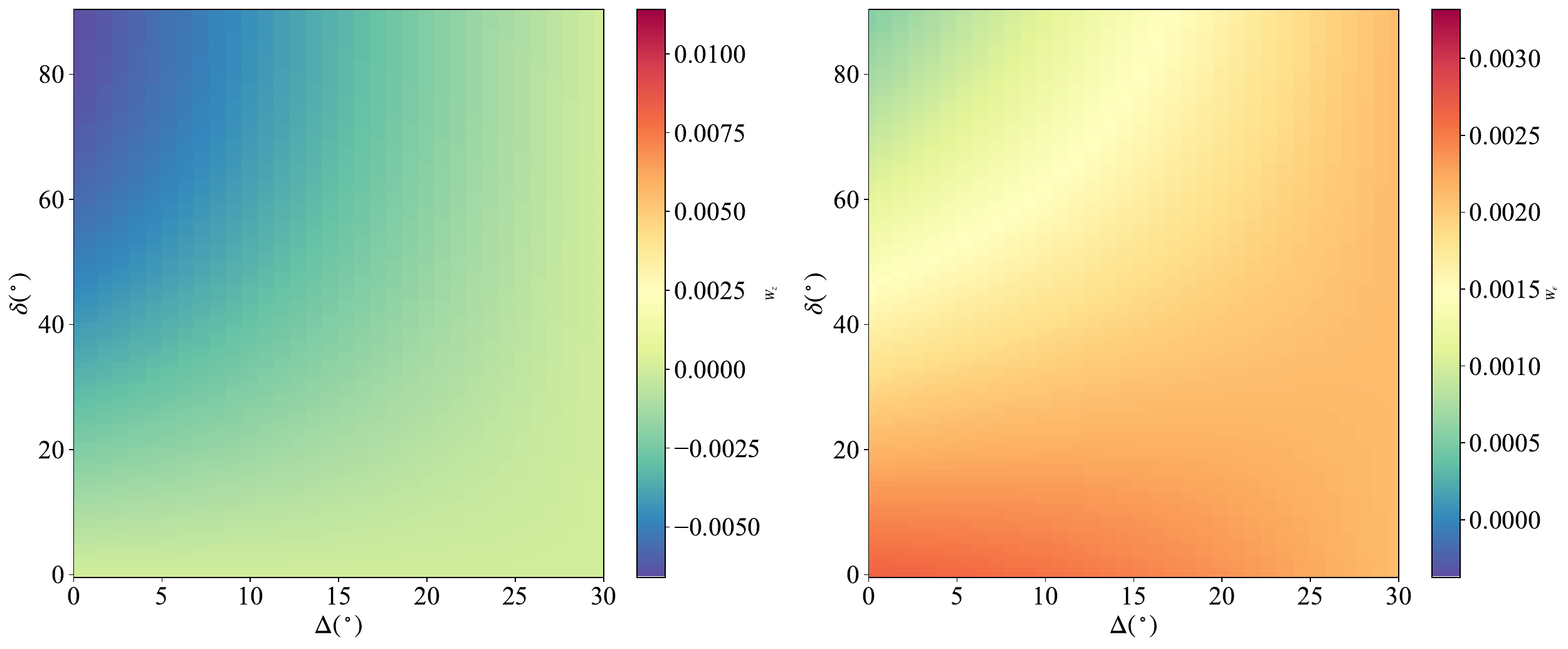}
    \caption{Spin component (left panel) and the obliquity component (right panel) of the CYORP coefficient as a function of $\Delta (< \alpha)$ and $\delta$, for $K = 2.65\rm W/m/K$. Here, $d/D_0 = 0.16$ and $\alpha = \epsilon = \pi/6$.}
    \label{fig:delta_Delta}
\end{figure*}

\begin{figure}
    \centering
    \includegraphics[width = 0.5 \textwidth]{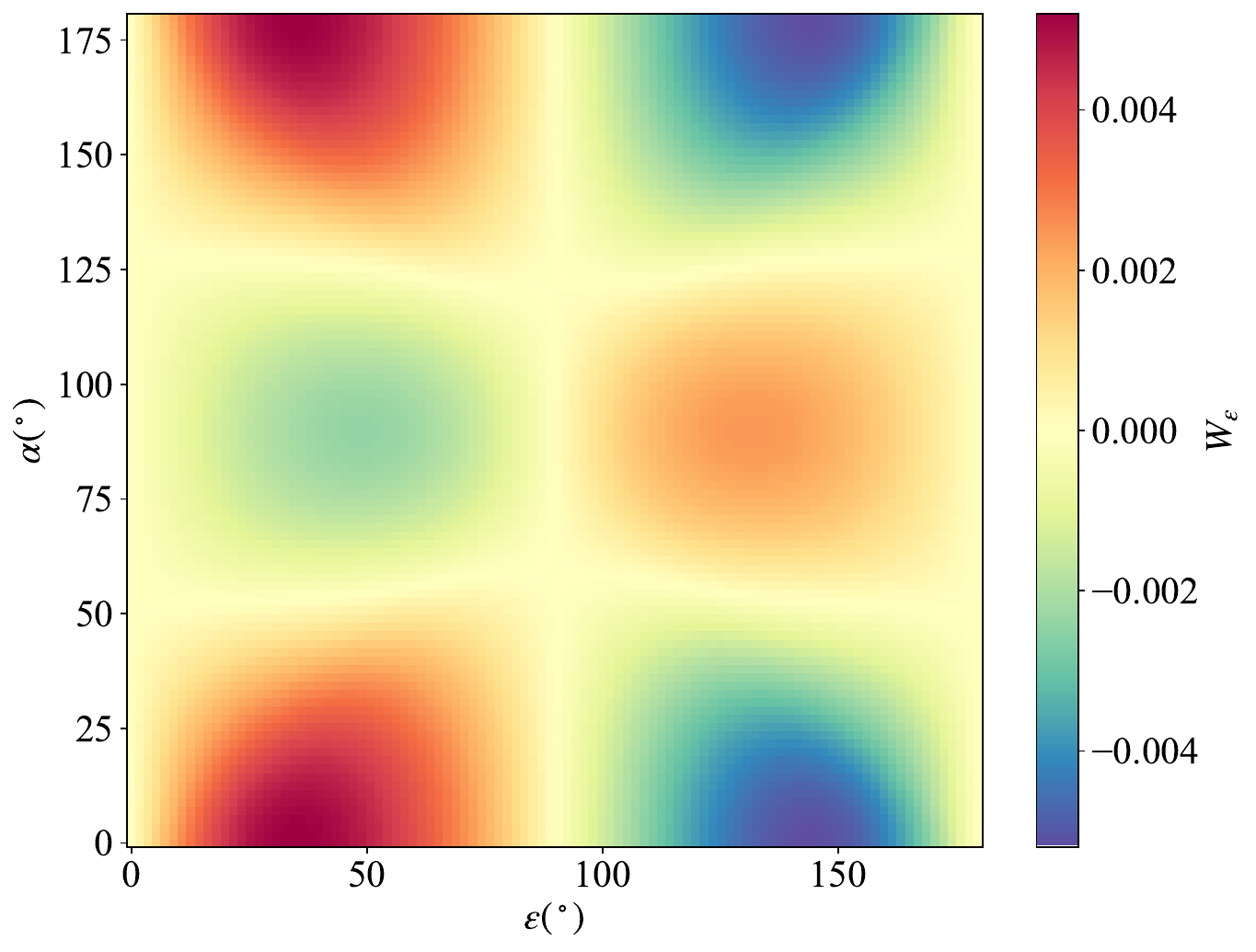}
    \caption{Obliquity component (right panel) of the CYORP coefficient as a function of the asteroid obliquity $\epsilon$ and the crater colatitude $\alpha$ for $K = 2.65 \rm W/m/K$. Here, $d/D_0 = 0.16$ and $\delta = \Delta = 0$. }
    \label{fig:zero_delta}
\end{figure}

\subsection{Irregularity $\delta$ and $\Delta$}

The angular parameters $\delta$ and $\Delta$ are used to describe the irregularity of the asteroid, where $\delta = 0$ and $\Delta = 0$ correspond to a perfect sphere. We explore the CYORP torque as a function of $\delta$ and $\Delta$ with fixed asteroid obliquity and crater colatitude of $\pi/4$. The results are presented in Fig.~\ref{fig:delta_Delta}. We can see that for the spin component, $\delta$ controls the torque magnitude while $\Delta$ controls the torque direction.

\citet{Zhou2022} demonstrates that the CYORP torque vanishes for $\delta = 0$. However, in the presence of finite thermal inertia, the obliquity component of the CYORP torque arises while the spin component remains negligible. Figure \ref{fig:zero_delta} illustrates the variation of the CYORP obliquity component with the asteroid obliquity and the crater colatitude when $\delta$ and $\Delta$ are both zero. The CYORP torque has a tendency to lead the asteroid obliquity to $90\circ$ when the crater is near the poles, while it leads to an asymptotic obliquity of $0^\circ$ or $180^\circ$ when the crater is near the equator.

\section{Application of the CYORP effect on a real asteroid}
\label{sec6}

\subsection{Roughness}
The surface roughness of asteroids is produced by several processes, including micrometeorite impacts, thermal fatigue, ejecta, or outgassing. It was found that the YORP torque is extremely sensitive to the small-scale surface structures \citep{Statler2009,Breiter2009}. The microscopic beaming effect of regolith grain-size-scale roughness (<1~mm) was shown ---using the Hapke reflectance and emissivity model \citep{Breiter2011}--- to have a marginal influence on the YORP effect. The transverse heat conduction across thermal skin depth ($\sim 1~$cm) causes an asymmetric thermal emission of the east and west sides of a boulder, giving rise to a systematic positive YORP torque \citep{Golubov12, Golubov2022}. The importance of the various self-modification effects of a concave feature on the surface was considered gradually and numerical approaches were taken to study it \citep{Statler2009, Rozitis2012, Rozitis2013a}. It was found that the concave feature of surface roughness could dampen the YORP torque by tens of percent. While the pioneering studies by \citet{Rozitis2012} shed light on the effects of roughness, the computational expense and difficulty in studying the functional dependence means that there are severe limitations to the numerical method. In contrast, the analytical method that we introduce in the present study, and its computational efficiency, simplify the application of roughness-induced YORP effects to real asteroids or binary systems. Given the objective of our semi-analytical method to provide a rapid assessment of the impact of surface roughness, it is particularly well-suited for models of asteroids with rough
convex shapes derived from light-curve observations. However, when dealing with high-resolution shape models, especially those that possess a resolution of a few centimetres (the scale of the thermal skin), a 3D thermal model becomes essential for accurate calculations, {{owing to the presence of the tangential YORP (TYORP) effect, which accounts for the transverse heat conduction inside a boulder.}}

\begin{figure*}
    \centering
    \includegraphics[width= \textwidth]{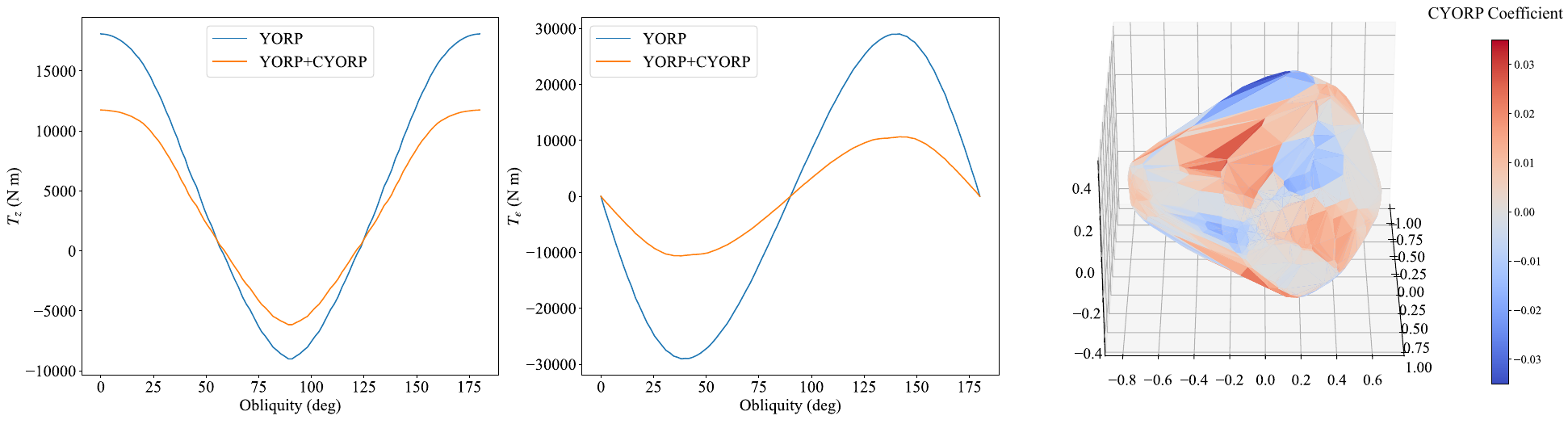}
    \caption{ YORP torque damped by the CYORP effect in the case of asteroid (272) Antonia. The spin component and the obliquity component are shown in the left and middle panels, respectively. The right panel shows the CYORP coefficient distribution over the asteroid's surface.}
    \label{fig:CYORP_damping}
\end{figure*}

For illustrative purposes, to demonstrate the application of our method, we randomly selected the main belt asteroid (272) Antonia as an example. This choice is representative of the majority of asteroids lacking detailed information obtained through in situ observations. We used the shape model obtained from \citet{Hanuvs2013}. To optimise the performance of our model, we assume a relatively high thermal conductivity of 1~W/m~K. We uniformly distributed the roughness across the asteroid's surface and investigated the total YORP torque (NYORP + CYORP). {{To do so, we input the normal vector and position vector of surface elements in the shape model to our CYORP model to obtain the CYORP torque coefficient of each surface element. We then used the area of each surface element to calculate the CYORP torque and sum up all CYORP torques generated by all surface elements. The depth-to-diameter is assumed to be 0.5, following the assumption made by \citet{Rozitis2012}.}} This has been compared with the sole NYORP torque. The result is depicted in Fig.~\ref{fig:CYORP_damping}. Our findings confirm that the roughness-induced CYORP torque damps the normal YORP torque. Specifically, for the asteroid (272) Antonia, the spin component of the torque experiences a damping effect of approximately 35\%, while the obliquity component is damped by approximately 64\%. 

\begin{figure*}
    \centering
    \includegraphics[width= \textwidth]{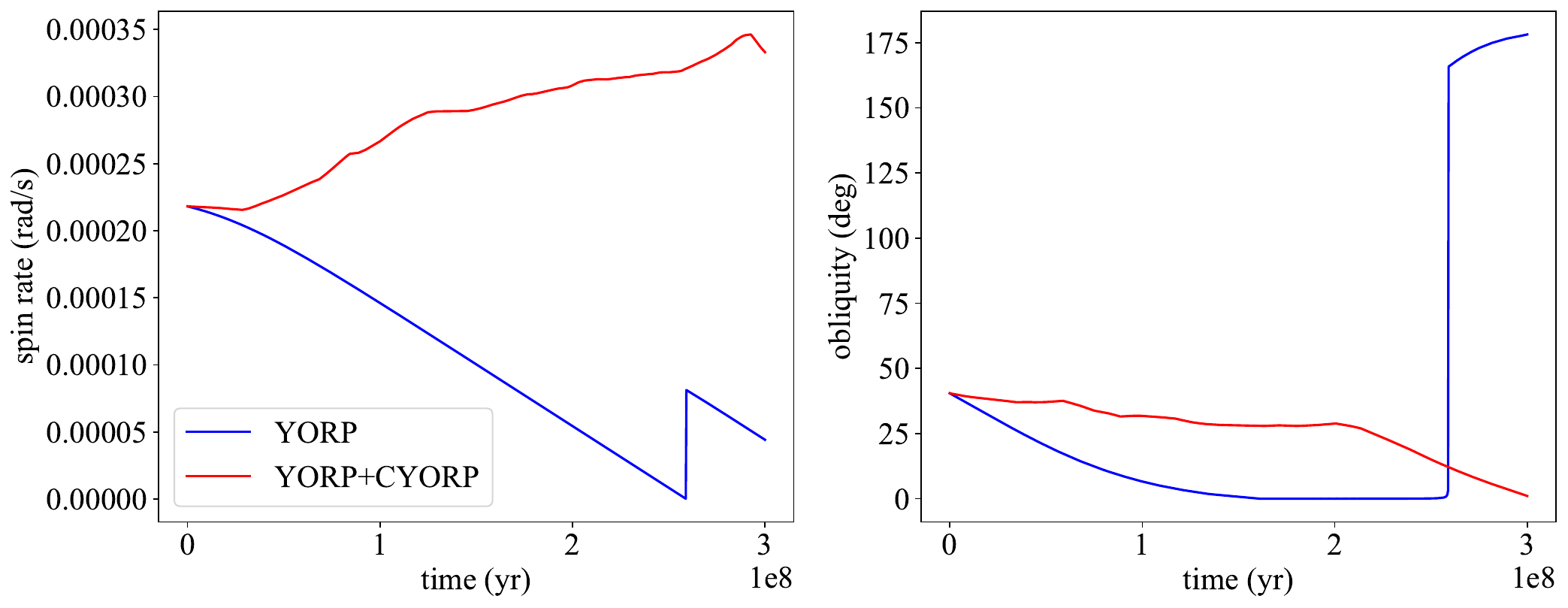}
    \caption{Evolution of the spin rate (left panel) and the obliquity (right panel) of a 10~km synthetic asteroid. In the presence of the static YORP torque (blue line), the asteroid gradually decelerates until it reaches a quasi-non-rotational state. Subsequently, {{we impose a new rotational state on the asteroid by assigning random values of rotational speed and obliquity. Conversely, when incorporating the CYORP torque (red line), the asteroid follows a different path, exhibiting random fluctuations in its spin rate due to the occurrence of impacts, creating new craters that lead to changes in the CYORP torque. As a result, the asteroid experiences intermittent transitions between spin up and spin down.}}}
    \label{fig:Exm_spin_evolution}
\end{figure*}

\subsection{Rotational evolution}
The rotational dynamics of asteroids are primarily governed by two key processes: collisions and the YORP effect. The timescale for reorientation of the spin axis resulting from angular momentum transfer during a collision can be expressed as \citep{Farinella1998}
\begin{equation}
    \tau_{\rm imp,re-ori} = 746 \left( {R_{\rm ast} \over 1 {\rm km}} \right)^{4/3} \left( {\omega \over 3 \times 10^{-4} {\rm s^{-1}} } \right)^{5/6} \rm Myr.
\end{equation}
On the other hand, the typical timescale associated with the YORP effect is approximately given by: 
\begin{equation}
    \tau_{\rm YORP} \sim 4 \left( {R_{\rm ast} \over 1 {\rm km}} \right)^2 \left( {\omega \over 3 \times 10^{-4} {\rm s^{-1}} } \right) \rm Myr.
\end{equation}
Clearly, the YORP timescale is shorter than $\tau_{\rm imp,re-ori}$, although its specific value exhibits considerable variation across different asteroids. Consequently, it is widely accepted that the YORP effect primarily governs the rotational evolution of small objects, while collisions play a dominant role in larger objects. {{The classic static YORP model ---which assumes an invariable YORP torque until the asteroid spins up to disruption or spins down to a quasi-static rotation--- has been used to study the long-term rotational evolution of asteroids \citep{Rubincam00, Pravec2005, Bottke2015}.  }}

A more intricate model, referred to as the `stochastic YORP model' \citep{Bottke2015}, takes into account the resetting of the YORP torque caused by collisions, which arises from the high sensitivity of the YORP effect to surface morphology. Although a suggested timescale of 1 Myr has been proposed for YORP reset \citep{Bottke2015}, a quantitative assessment of the torque changes resulting from craters is yet to be explored. The CYORP effect offers a powerful tool for investigating the stochastic YORP effect. While a comprehensive examination of the stochastic YORP effect is beyond the scope of this paper, we present an example of integrating the CYORP effect into the static YORP model.

In our simulation, a random YORP coefficient is assigned within the range of -0.005 to 0.005, with the coefficient's sign matching that of the torque. The CYORP torque is introduced specifically when a collision event takes place. The timescale for the impact by an asteroid with the size $R_{\rm imp}$ is
\begin{equation}
\label{eq:tau_imp}
    \tau_{{\rm imp}} = \frac{1}{P_{\rm i} \pi R_{\rm Antonia}^2 \Delta N(R>R_{\rm imp})},
\end{equation}
where 
\begin{equation}
    N(R > R_{\rm imp}) = C_R \left( \frac{R_{\rm imp}}{1~{\rm km}}  \right)^{-b_R}.
\end{equation}
Here, $P_{\rm i} = 2.85 \times 10^{-18} \rm km^{-2} yr^{-1}$ is the intrinsic collision probability, $C_R = 6 \times 10^5$, and $b_R = 2.2$ \citep{Holsapple2022}. In the strength regime formulation, the crater produced by an impactor with the size of $R_{\rm imp}$ has a size of
\begin{equation}
\label{eq:R_crater_strength}
    R_{\rm crater} = 1.306 R_{\rm imp} \left( { \rho_{\rm imp} \over \rho_{\rm ast}  }^{0.4} \right) \left( {Y\over \rho v_{\rm imp}} \right)^{-0.205},
\end{equation}
with $Y = 100$~Pa and $v_{\rm imp} = 5.3$~km/s. In the gravity regime, 
\begin{equation}
    R_{\rm crater} = 0.59 \left( {\rho_{\rm ast} \over m_{\rm imp} } \right)^{-1/3}  \left( {\rho_{\rm ast} \over \rho_{\rm imp} }\right)^{0.00138} \left( {g_{\rm ast} R_{\rm ast} \over v_{\rm imp}} \right)^{-0.17},
\end{equation}
where $g_{\rm ast}$ is the surface gravity of the asteroid. {{In each time step ($\sim 10^3$ years), we calculate the numbers of impact craters of different sizes, according to Eq.~\ref{eq:tau_imp} and Eq.~\ref{eq:R_crater_strength}. We then assign each crater with a random surface element of the polyhedron model of the asteroid Antonia, after which we can calculate the CYORP torque. Finally, we add the CYORP torque to the normal YORP torque directly calculated from the shape model in order to obtain the total YORP torque.
The spin rate evolves following
\begin{equation}
    \dot \omega = {T_z \over I},
\end{equation}
with $I$ being the maximum moment of inertia and $T_z$ being the torque component that is along a spin vector.
The obliquity evolves according to
\begin{equation}
    \dot \epsilon = {T_\epsilon \over I \omega}
,\end{equation}
where $T_\epsilon$ is the torque component that changes the obliquity.}}

There exist two possible end states in a YORP cycle: either the asteroid's rotation slows down until it reaches a quasi-non-rotational state, or it accelerates to the spin threshold for shape change or disruption with a period of approximately 2.2 hours. Upon completing a YORP cycle, the asteroid's rotational state is updated by assigning a new random rotational speed and obliquity. The impact of introducing CYORP torques can be observed in the evolution of a 10 km asteroid, as depicted in Fig.~\ref{fig:Exm_spin_evolution}. Notably, significant differences arise when considering the inclusion of CYORP torques.

Nonetheless, the rotational evolution of asteroids currently lacks a standardised model. Some models propose that after spinning down to a non-rotational state, the asteroid's spin rate is assigned a new random value within a specified range \citep{, Hanuvs2011, Bottke2015}, while some assume it continues to spin up under the YORP effect \citep{Pravec2008,Marzari2020}. By selecting an initial spin rate for a new rotational state, \citet{Holsapple2022} reproduces the spin evolution without the YORP effect. Hence, rather than attempting to address the entire complexity of the problem, our objective in this study is to present an illustrative example of the interaction between  CYORP and the conventional YORP effect. Furthermore, we underscore the significance of the CYORP effect in the long-term rotational evolution of asteroids. A comprehensive investigation of the rotational evolution of asteroid groups is left for future research.

\section{Summary and conclusions}
\label{sec7}

The YORP effect is a thermal torque produced by radiation from the irregular surface of the asteroid. It has been demonstrated that this effect is highly sensitive to surface topology \citep{Statler2009, Breiter2009}, including small-scale roughness \citep{Rozitis2012}, boulders \citep{Golubov12}, and craters \citep{Zhou2022}. In this study, we developed a semi-analytical model for calculating the temperature field of a crater, which accounts for the effects of self-sheltering, self-radiation, self-scattering, and non-zero thermal conductivity. Using this model, we investigated the crater-induced YORP (CYORP) effect in a computationally efficient manner (about three orders of magnitude faster than the numerical method), allowing for a comprehensive exploration of the functional dependence of the CYORP effect and its incorporation into the rotational and orbital evolution of asteroids. The main results and conclusions of this study can be summarised as follows.

Our semi-analytical model for the CYORP effect is valid in the high-thermal-conductivity regime ($K > 0.1 \rm W/m/K$). This suggests that the model is suitable for application to materials such as solid basalt and metal, which are usually beneath the regolith on asteroid surfaces but may be exposed to sunlight due to the formation of deep craters. 

The CYORP effect is significant when the crater depth-to-diameter ratio is greater than 0.05. The self-modification effects of a concave structure, including the self-sheltering effect, self-radiation effect, and self-scattering effect, are stronger with a higher depth-to-diameter ratio. For concave structures with a depth-to-diameter ratio of smaller than 0.05, the surface can be treated as a convex shape without introducing significant inaccuracies. The typical value of the CYORP coefficient for the spin component is 0.01, which is insensitive to the thermal parameter, while the obliquity component decreases from 0.004 for basalt to 0.001 for metal. For a z-axis symmetric shape (e.g. a spinning top shape), the spin component of the CYORP torque vanishes while the obliquity component survives, which implies that the spin acceleration of such symmetric shapes does not change significantly under the effect of crater formation.

Using our semi-analytical method, we confirm that the YORP torque can be damped by the surface roughness, which was first discovered by \citet{Rozitis2012}. The fast computation of our semi-analytical model allows us to consider more flexible configurations of surface roughness, such as a space-varying roughness distribution, roughness on components of binary asteroids, and so on.

The magnitude and directional change of the YORP coefficient at each impact are assessed for the first time using our CYORP model. While a complete investigation of the spin evolution of asteroids is left for future work, we show that rotational evolution can be severely affected by collisions.

\begin{acknowledgements}
 We acknowledge support from the Universit\'e C\^ote d'Azur. We thank Yun Zhang, Xiaoran Yan, and Marco Delbo for the useful discussion. Wen-Han Zhou would like to acknowledge the funding support from the Chinese Scholarship Council (No.\ 202110320014). Patrick Michel acknowledges funding support from the French space agency CNES and from the European Union's Horizon 2020 research and innovation program under grant agreement No.\ 870377 (project NEO-MAPP). 
\end{acknowledgements}

%
%
\bibliographystyle{aa} 
\bibliography{references}

\end{document}